\documentclass[11pt,oneside,a4paper]{IEEEtran}
\usepackage{multirow}
\usepackage{amsmath}
\usepackage{mathrsfs}
\usepackage{array}

\usepackage[dvipsnames]{xcolor}

\usepackage{tikz,pgfplots}
\usepgfplotslibrary{units}
\usetikzlibrary{positioning,shadows,shapes,fit,arrows,backgrounds}

\colorlet{figyellow}{yellow!40!white}
\colorlet{figred}{red!40!white}
\colorlet{figblue}{blue!20!white}
\colorlet{figgreen}{green!30!white}
\colorlet{figgray}{black!10!white}
\colorlet{figwhite}{white!10!white}

\usetikzlibrary{external}
  \DeclareGraphicsExtensions{.pdf,.jpeg,.png,.jpg}

\newcommand{\mytilde}{\raise.17ex\hbox{$\scriptstyle\mathtt{\sim}$}}

\newcommand{\dla}[1]{#1}

\begin{document}
\title{\dla{Why compensating fibre nonlinearity will never meet capacity demands}}

\author{Domani\c{c} Lavery,
	Robert~Maher,
        David~Millar,
        Alex~Alvarado,
        Seb~J.~Savory,
        and~Polina~Bayvel% <-this % stops a space
        
%\thanks{Manuscript received June 29, 2012; revised August 30, 2012, September 28, 2012; accepted September 28, 2012.}
\thanks{This work was made possible through financial support from EPSRC through project EP/J017582/1 UNLOC (Unlocking the capacity of optical communications).  D.~Lavery, R.~Maher, A.~Alvarado, S.~J.~Savory, and P.~Bayvel are with the Optical Networks Group, UCL Electronic and Electrical Engineering, Torrington Place, London, WC1E 7JE, UK (e-mail: d.lavery@ee.ucl.ac.uk). D.~Millar is with Mitsubishi Electric Research Laboratories, 201 Broadway, Cambridge, MA, 02139 USA (email: millar@merl.com).}}%
%\thanks{Color versions of one or more of these figures in this paper are available online at http://www.ieeexplore.ieee.org.}%

%\markboth{}%
%{Lavery \MakeLowercase{\textit{et al.}}: Approaching the Fundamental Limits of Classical Optical Communication Using Single Mode Fibre}

\maketitle

\begin{abstract}
Current research efforts are focussed on overcoming the apparent limits of communication in single mode optical fibre resulting from distortion due to fibre nonlinearity. It has been experimentally demonstrated that this Kerr nonlinearity limit is not a fundamental limit; thus it is pertinent to review where the fundamental limits of optical communications lie, and direct future research on this basis. This paper details recently presented results \cite{RSMeet}.  The work herein briefly reviews the intrinsic limits of optical communication over standard single mode optical fibre (SMF), and shows \dla{that the empirical limits of silica fibre power handling and transceiver design both introduce a practical upper bound to the capacity of communication using SMF}, on the order of 1 Pbit/s. \dla{Transmission rates exceeding 1 Pbit/s are shown to be possible, however, with currently available optical fibres, attempts to transmit beyond this rate by simply increasing optical power will lead to an asymptotically zero fractional increase in capacity}.
\end{abstract}

%\begin{IEEEkeywords}
%Optical access, ultra dense (UD), wavelength division multiplexing (WDM), passive optical network (PON), digital signal processing (DSP).
%\end{IEEEkeywords}

\section{Introduction}
\begin{figure}
\centering
\resizebox{1\columnwidth}{!}{%
\centerline{
\footnotesize{
\begin{tikzpicture}[plain/.style={align=center,execute at begin node=\setlength{\baselineskip}{2.5ex}}]
\node[plain] at (-145pt,0pt) {\textbf{(a)}};
% Tx
%\draw[thick,fill=figred,rounded corners=3pt] (-160pt,-15pt) rectangle (-110pt,15pt);
%\node[plain] at (-135pt,5pt) {Optical};
%\node[plain] at (-135pt,-5pt) {Transmitter};
\draw[thick,fill=figred,rounded corners=3pt] (-128pt,-15pt) rectangle (-110pt,15pt);
\node[plain] at (-119pt,0pt) {Tx};
% Lines
\draw[thick,-] (-110pt,0pt) -- (-50pt,0pt); % Long Line
\draw[thick,->] (-25pt,0pt) -- (35pt,0pt); % Long Line
%
% Span
\node[plain,anchor=south] at (-88pt,20pt) {SMF};
\draw[thick,-] (-91pt,10pt) circle (10pt);
\draw[thick,-] (-88pt,10pt) circle (10pt);
\draw[thick,-] (-85pt,10pt) circle (10pt);
%
% More lines
\draw[loosely dotted] (-55pt,0pt) -- (-20pt,0pt); % Long Line
\draw[thick,-,fill=white] (-70pt,-10pt) -- (-55pt,0pt) -- (-70pt,10pt) -- (-70pt,-10pt);%EDFA
% Span
\node[plain,anchor=south] at (-8pt,20pt) {SMF};
\draw[thick,-] (-11pt,10pt) circle (10pt);
\draw[thick,-] (-8pt,10pt) circle (10pt);
\draw[thick,-] (-5pt,10pt) circle (10pt);
\draw[thick,-,fill=white] (10pt,-10pt) -- (25pt,0pt) -- (10pt,10pt) -- (10pt,-10pt);%EDFA
% Rx
%\node[plain] at (12pt,-15pt) {};
\draw[thick,fill=figblue,rounded corners=3pt] (35pt,-15pt) rectangle (53pt,15pt);
\node[plain] at (44pt,0pt) {Rx};
%\node[plain] at (60pt,5pt) {Optical};
%\node[plain] at (60pt,-5pt) {Receiver};
% Grid
%\draw[step=5pt,gray,very thin] (-250pt,-50pt) grid (250pt,50pt);
%\foreach \x in {-250,-240,...,250} \draw (\x pt,-50pt) node[anchor=east,rotate=90] {\tiny $\x$};
%\foreach \y in {-50,-40,...,50} \draw (-250pt, \y pt) node[anchor=east] {\tiny $\y$};
\end{tikzpicture}}}
}
% Next
\resizebox{1\columnwidth}{!}{%
\centerline{
\footnotesize{
\begin{tikzpicture}[plain/.style={align=center,execute at begin node=\setlength{\baselineskip}{2.5ex}}]
\node[plain] at (-145pt,0pt) {\textbf{(b)}};
% Tx
\draw[thick,fill=figred,rounded corners=3pt] (-128pt,-15pt) rectangle (-110pt,15pt);
\node[plain] at (-119pt,0pt) {Tx};
% Lines
\draw[thick,-] (-110pt,0pt) -- (-50pt,0pt); % Long Line
\draw[thick,-] (-25pt,0pt) -- (35pt,0pt); % Long Line
%
% Span
\node[plain,anchor=south] at (-88pt,20pt) {SMF};
\draw[thick,-] (-91pt,10pt) circle (10pt);
\draw[thick,-] (-88pt,10pt) circle (10pt);
\draw[thick,-] (-85pt,10pt) circle (10pt);
%
% OEO
\draw[thick,fill=figblue,rounded corners=3pt] (-62pt,-15pt) rectangle (-28pt,15pt);
\draw[thick,fill=figred,rounded corners=3pt] (-32pt,-15pt) rectangle (-12pt,15pt);
\draw[thick,fill=figwhite,rounded corners=3pt] (-46pt,-15pt) rectangle (-28pt,15pt);
\node[plain] at (-54pt,0pt) {Rx};
\node[plain,rotate=90] at (-37pt,0pt) {Elec.};
\node[plain] at (-20pt,0pt) {Tx};
\node[plain,anchor=south] at (-37pt,15pt) {OEO};
%
%\draw[thick,-,fill=white] (-70pt,-10pt) -- (-55pt,0pt) -- (-70pt,10pt) -- (-70pt,-10pt);%EDFA
% Span
\node[plain,anchor=south] at (13pt,20pt) {SMF};
\draw[thick,-] (10pt,10pt) circle (10pt);
\draw[thick,-] (13pt,10pt) circle (10pt);
\draw[thick,-] (16pt,10pt) circle (10pt);
%
% More lines
\draw[loosely dotted] (0pt,0pt) -- (53pt,0pt); % Long Line
% Grid
%\draw[step=5pt,gray,very thin] (-250pt,-50pt) grid (250pt,50pt);
%\foreach \x in {-250,-240,...,250} \draw (\x pt,-50pt) node[anchor=east,rotate=90] {\tiny $\x$};
%\foreach \y in {-50,-40,...,50} \draw (-250pt, \y pt) node[anchor=east] {\tiny $\y$};
\end{tikzpicture}}}
}
% Next
\resizebox{1\columnwidth}{!}{%
\centerline{
\footnotesize{
\begin{tikzpicture}[plain/.style={align=center,execute at begin node=\setlength{\baselineskip}{2.5ex}}]
\node[plain] at (-145pt,0pt) {\textbf{(c)}};
\draw[thick,fill=figred,rounded corners=3pt] (-128pt,-15pt) rectangle (-110pt,15pt);
\node[plain] at (-119pt,0pt) {Tx};
%\node[plain] at (-95pt,-5pt) {Transmitter};
\draw[thick,fill=figgray,dashed,rounded corners=3pt] (-5pt,-15pt) rectangle (27pt,15pt);
\node[plain,anchor=south] at (-35pt,20pt) {SMF};
\draw[thick,-] (-38pt,10pt) circle (10pt);
\draw[thick,-] (-35pt,10pt) circle (10pt);
\draw[thick,-] (-32pt,10pt) circle (10pt);
\draw[thick,->] (-110pt,0pt) -- (35pt,0pt); % Long Line
\draw[thick,-,fill=white] (2pt,-10pt) -- (17pt,0pt) -- (2pt,10pt) -- (2pt,-10pt);
\node[plain] at (12pt,-15pt) {};
\draw[thick,fill=figblue,rounded corners=3pt] (35pt,-15pt) rectangle (53pt,15pt);
\node[plain] at (44pt,0pt) {Rx};
%\node[plain] at (60pt,-5pt) {Receiver};
% Grid
%\draw[step=5pt,gray,very thin] (-250pt,-50pt) grid (250pt,50pt);
%\foreach \x in {-250,-240,...,250} \draw (\x pt,-50pt) node[anchor=east,rotate=90] {\tiny $\x$};
%\foreach \y in {-50,-40,...,50} \draw (-250pt, \y pt) node[anchor=east] {\tiny $\y$};
\end{tikzpicture}}}
}
\caption{\dla{(a) A multi-span transmission system, (b) a multi-span transmission system, which uses optical-electrical-optical regeneration after each span, and (c) the transmission system under consideration, which is a subset of transmission system (b).  As noted in \cite{Kikuchi2008}, the signal-to-noise ratio limit of this system is equivalent with or without an optical amplifier \dla{prior to the receiver}.}}
\label{fig:config}
\end{figure}
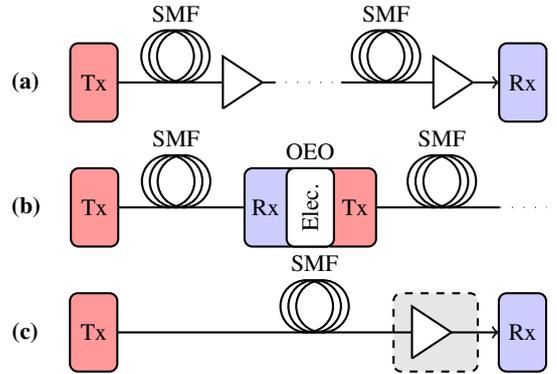

The last two decades have seen extraordinary increases in both the demonstrated information carrying capability of single mode optical fibres (SMF) and the efficiency with which the optical spectrum is used, which raises the following questions: i) Is there a maximum rate at which information can be transmitted in an optical fibre, and ii) If there is a limit, can it be quantified?  \dla{To answer these questions, it is tempting to use the approach taken in, e.g., \cite{Essiambre2010,Mitra2001}, which studies the transmission system in Fig.~\ref{fig:config}(a).  Here,} two distinct noise sources are considered: amplified spontaneous emission (ASE) noise from optical amplifiers, and interference due to the power-dependent nonlinear signal distortion.  Although the nonlinear distortion is deterministic, these works assume that the nonlinear distortion is uncompensated\footnote{\dla{That is, the signal-signal nonlinear interaction is determininistic, although the signal-ASE nonlinear interactions are not. These works assume that neither interaction is compensated.}}, and, therefore, both the ASE and nonlinear distortions can be treated as additive white Gaussian noise (AWGN). \dla{The assumption of uncompensated nonlinear interference results in an achievable information rate (AIR) in these models, but not the highest AIR; that is, these works lower bound optical fibre capacity}.  Indeed, these rates have been \dla{experimentally} shown to be achievable, and these works (and derivative works which also employ perturbative nonlinear noise models) are frequently used to accurately estimate the performance of optical transmission systems.  However, by mitigating the nonlinear interference in the receiver, it is now trivial to demonstrate \dla{transmission throughputs} beyond these lower bounds \cite{Maher2015}.

%\begin{figure*}[!tb]
%  \centering
%    \includegraphics[width=0.7\textwidth]{Figures/Config}
%\caption{Simulated link for investigating the performance of fiber nonlinearity compensation, where the DSP is divided between transmitter and receiver.}
%\label{fig:config}
%\end{figure*}

More recently, Kramer, Yousefi and Kschischang \dla{\cite{Kramer2015, Mansoor2015}} \dla{showed} that an upper bound for this maximum achievable rate is simply the limit of communication in a linear AWGN channel with ASE noise, only.  That is, the nonlinear interference in SMF cannot be used to overcome the limits to optical communication imposed by AWGN.  One could, of course, use nonlinear elements to increase the achievable information rate beyond the limit imposed by ASE noise, and a simple example of such a nonlinear element is an optical-electrical-optical (OEO) regenerator\dla{, as in Fig.~\ref{fig:config}(b)}.

The use of such idealised regenerators reduces the performance analysis of a multi-span system to that of a single fibre span \dla{(Fig.~\ref{fig:config}(c))}.  \dla{It has been shown that, in single span transmission systems, a maximum likelihood receiver can, in some circumstances, ideally detect a nonlinearly distorted signal, returning the system performance to the AWGN channel limit \cite{Liga2015}}.  In optical communication systems, shot noise -- or, photon noise -- introduces noise in all classical receivers \dla{(i.e., a receiver which does not take account of quantum effects)}.  Thus, the data processing inequality, which states that the information content of a signal cannot increase by any local physical operation, inherently limits the maximum achievable rate for any optical communication system to the limit imposed by the noise introduced by a receiver.  In this paper, we further analyse this observation using the model in Fig.~\ref{fig:config}(c) under the assumption of an infinitesimally short length of fibre, and examine the implications for future optical communications systems.

\section{Revisiting the Standard Quantum Limit}
\begin{figure*}
\centering
\sffamily{
\begin{tikzpicture}
\begin{loglogaxis}[width=0.90\textwidth,
	height=0.4\textwidth,,
	every axis/.append style={font=\small},
	ylabel={Capacity Upper Bound}, 
	y unit=Tb/s,
	xlabel= {Received Optical Power},
	x unit=W,
	ymin=1,
	ymax=2e4,
	grid=both, 	
	legend style={legend pos=south east,font=\scriptsize,legend cell align=left},
	xmin=1e-7,
	xmax=1e30,
]
\addplot+[color=red,dashed,mark=none,mark size=2,mark options={fill=black}, thick] file {plots/Fig2_fullBW.txt};
\addplot+[color=blue,solid,mark=none,mark size=2,mark options={fill=black}, thick] file {plots/Fig2_15THz.txt};
\addplot+[color=OliveGreen,dashdotted,mark=none,mark size=2,mark options={fill=black}, thick] file {plots/Fig2_11.2THz.txt};
\addplot+[only marks,color=black,solid,mark=*,mark size=3,mark options={fill=black}] file {plots/Fig2_Sano.txt};
%\addplot [only marks,mark=*] coordinates { (0,0) };
\addplot+[color=black,solid,mark=none,mark size=2,mark options={fill=black},very thick] file {plots/Fig2_Sun.txt};
\addplot [color=black,solid,mark=none, thick] coordinates { (1e-7,1e3) (1e30,1e3)};
\addplot [color=black,solid,mark=none, thick] coordinates { (1e-7,1e4) (1e30,1e4)};

\node at (axis cs:8e-8,9.4e2) [anchor=south west] {\footnotesize 1~Pb/s};
\node at (axis cs:8e-8,9.4e3) [anchor=south west] {\footnotesize 10~Pb/s};

\legend{O-U band capacity (50 THz), C-L band capacity (15 THz), 11.2~THz capacity, Record throughput 102.3 Tb/s (11.2~THz), Solar luminosity};
\end{loglogaxis}
\end{tikzpicture}
}
\caption{Capacity upper bounds \dla{for 11.2, 15 and 50 THz channels.  The marker indicates an experimentally demonstrated result, achieved in an 11.2~THz bandwidth} (Sano \textit{et al}. \cite{Sano2012}).  Transmission in all cases assumes the use of both orthogonal polarisation states supported by `single mode' optical fibre.  The thick vertical reference line indicates the solar luminosity \cite{Pradhan2011}.}
\label{fig:results}
\end{figure*}
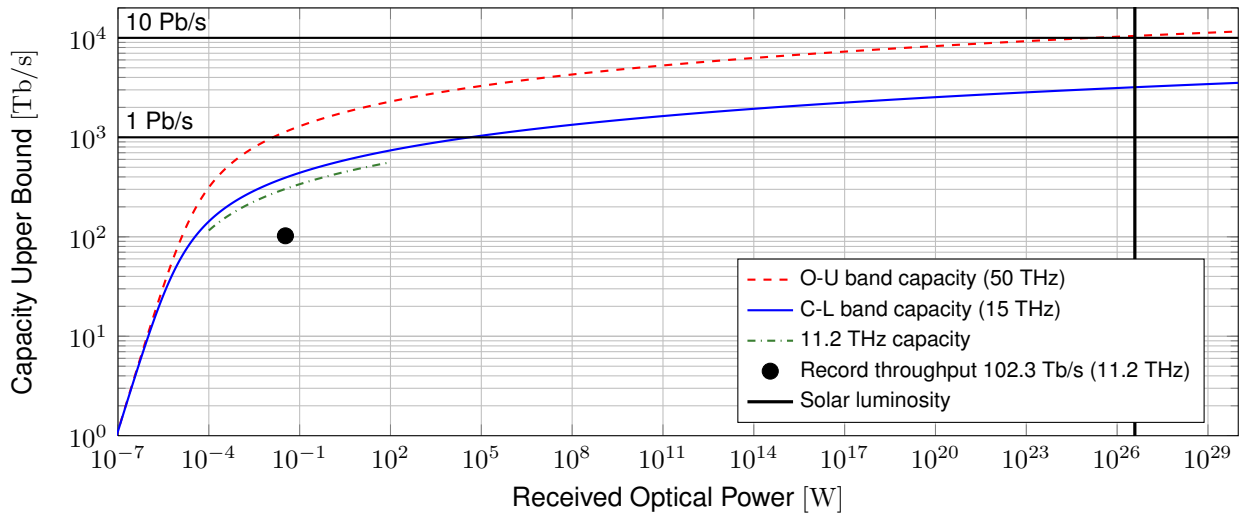

% Shannon limit
Claude Shannon, in his seminal work, defined capacity as the maximum rate at which information can be transmitted can be transmitted through a channel, with an arbitrarily low error rate \cite{Shannon1948}.  A great deal of work was undertaken between the 1960s and late 1980s investigating the physical limits of capacity in bosonic\footnote{In this context, this refers to communication using photons.} communication channels.  The classical (as opposed to quantum) limits of communication are considered herein, as summarised in the work of Caves and Drummond \cite{Caves1994}.  In particular, we focus on the `standard quantum limit', which is itself derived from the classical Shannon capacity limit for an AWGN channel.

The capacity of an AWGN channel under an average power constraint is $C=B\log_2(1+\gamma_s)$, where $B$ is the signal bandwidth and $\gamma_s$ is the signal-to-noise ratio (SNR).  Observing that shot noise follows a Gaussian distribution for large numbers of incident photons, it is possible to define a meaningful SNR for a shot noise limited system.  This is given by $\gamma_s=P_s/h\nu{}B$ where $P_s$ is the received signal power, $h$ is Planck's constant \dla{\cite{Kikuchi2008}}, and $\nu$ is the carrier frequency.  The combination of these formulae is the standard quantum limit\footnote{A note of caution here, that as the bandwidth tends to infinity, or when the optical power tends towards zero, this formula loses physical meaning, as the quantum nature of light must be considered.}
\begin{equation}
\begin{aligned}
C=B\log_2{\left(1+\frac{P_s}{h\nu{B}}\right)}.
\label{eq:sql}
\end{aligned}
\end{equation}
This master formula is used for the results in the following section, and is a function of both bandwidth and power.  The following results consider linear transmission of a dual polarisation signal, and this can therefore be included by simply doubling the available bandwidth, without loss of generality.%  As the quantity under consideration is the capacity of an AWGN channel, the assumptions on modulation format and coding are as described in \cite{Shannon1948}.

\section{Capacity Upper Bounds}
The curves in Fig.~\ref{fig:results} show the capacity, $C$, versus received optical power, $P_s$, for dual polarisation optical signals with a finite bandwidth, $B$. The bandwidths considered are the combined C- and L-band (15~THz), and the combined O-, E-, S-, C-, L- and U-bands (50~THz).  The latter curve is approximately the full fibre bandwidth, limited by the single mode cut-off at low wavelengths, and parasitic infrared absorption at the high wavelengths.

To place some perspective on these capacity curves, we consider the recently demonstrated experimental results from Sano~\textit{et~al.} \cite{Sano2012} using 11.2~THz bandwidth from the C- and L-bands, which achieved a throughput of 102.3~Tbit/s (over 240~km optical fibre) using an aggregate optical signal power of 33.9~mW\footnote{\dla{This experimental demonstration used backward pumped Raman amplification, which applied distributed gain to the signal over each span.} Therefore, when quoting the achieved transmission rates in Fig.~\ref{fig:results}, the launched signal power is used.  This potentially overestimates the equivalent received optical power but, given the logarithmic dependence of capacity on power, this correction factor can be considered small.}.  For this optical power, the standard quantum limit \dla{in a 15~THz bandwidth} indicates a capacity upper bound of 393.1~Tbit/s; less than a factor of four greater than the experimental demonstration.  \dla{For the fraction of the C- and L-band used in this work (11.2~THz), the experimental demonstration is less than a factor of 3 from this theoretical limit, as shown in Fig~\ref{fig:results}}.  Due to the logarithmic dependence of capacity on power, the fractional increase in capacity for each doubling of optical power tends asymptotically to zero.  In other words, there are diminishing returns when increasing optical power to increase capacity.  In the same bandwidth (15~THz), the minimum received power required to achieve a capacity of 100~Tbit/s is less than 1~mW.  However, to achieve a factor of 10 increase in capacity (1~Pbit/s) in the same bandwidth, the optical power requirement is approximately $10^{5}$~W.

As noted in \cite{ErikRS}, and indicated above, any optical power restriction, whether practical or fundamental, will ultimately bound the optical fibre capacity. There exist empirical limits to optical fibre communication, such as fibre fuse \cite{Kashyap2013}, where high input optical powers cause localised heating of defects in silica, leading to catastrophic and unrecoverable failure of an optical fibre. This issue has previously been noted \cite{NewBreeze2013}, and research into the power handling limits of optical fibres is ongoing.  Strikingly, however, increasing the optical fibre power handling by several orders of magnitude (from watts to kilowatts) appears unlikely to significantly impact on this capacity upper bound.  One must then ask how much optical power is justifiable to seek future increases in optical fibre capacity.

Alternatively, increasing bandwidth approximately scales the capacity linearly \dla{(see discussion in appendix)}.  Combined C- and L-band transmission requires prohibitively high optical powers to achieve 1~Pbit/s, however transmission using the full fibre bandwidth (50~THz) achieves 1~Pbit/s with a minimum required optical power of approximately 10~mW; eminently achievable.  To achieve a capacity one order of magnitude above this figure (i.e., 10~Pbit/s), would, again, require prohibitively high received optical powers ($1.8\times{}10^{25}$~W); in this scenario, an optical power approaching the luminosity of the sun \cite{Pradhan2011}.  As significant bandwidth scaling beyond this is \dla{infeasible} in SMF, we consider 1~Pbit/s to be a practical upper bound on the capacity of communication over \dla{SMF}.

\section{Conclusions}
There are a number of apparent limits to the capacity of optical communication over single mode optical fibre, but many of these limits are empirical, rather than fundamental.  In recent years, much research effort has been invested in overcoming distortions introduced due to the Kerr nonlinearity; and rightly so.  However, it must be recognised that the gains in transmission throughput that can be achieved by fibre nonlinearity mitigation are upper bounded not merely by the properties of the transmission medium, but by the nature of light itself.  Recent experimental demonstrations of throughput in a single fibre core are within a factor of four of the standard quantum limit for combined C- and L-band transmission \cite{Sano2012}, but the optical power required to significantly increase the throughput beyond this limit is prohibitive. To address this fundamental issue, there needs to be a paradigm shift in optical communications, exploring significantly wider bandwidths and parallel transmission channels.

\appendix
\dla{It is noted in the preceding text that, to a first approximation, the capacity increases linearly with bandwidth.  The purpose of this appendix is to justify that assertion, and identify the regime in which this approximation is valid.  Consider Eq.~\eqref{eq:sql}, which is a function of bandwidth, $B$, appearing both inside and outside the logarithm. To evaluate the impact of signal bandwidth on capacity, one can take the derivative of Eq.~\eqref{eq:sql} with respect to bandwidth, which, using the product rule, evaluates as follows
\begin{equation}
\begin{aligned}
\frac{\partial{}C}{\partial{}B}=&B\frac{\partial{}}{\partial{}B}\left(\log_2{\left(1+\frac{P_s}{h\nu{B}}\right)}\right)\\&+\log_2{\left(1+\frac{P_s}{h\nu{B}}\right)}.
\label{eq:derivative_sql1}
\end{aligned}
\end{equation}
In the limit that $\left({P_s}/{h\nu{B}}\right)\gg1$, this becomes
\begin{equation}
\begin{aligned}
\frac{\partial{}C}{\partial{}B}=\log_2{\left(\frac{P_s}{h\nu{B}}\right)}-\frac{1}{\log_e(2)}.
\label{eq:derivative_sq2}
\end{aligned}
\end{equation}
Evaluating Eq.~\eqref{eq:derivative_sq2}, it can be seen that, although the gradient changes rapidly for small values of B, over the bandwidth considered in this work (between the 15~THz and 50~THz) this function can be approximated as stationary.
}

%
%\bibliography{summary}{}
%\bibliographystyle{IEEEtran}
%
% Generated by IEEEtran.bst, version: 1.13 (2008/09/30)

\end{document}